\documentclass[twocolumn]{article}%
\usepackage{amsmath}
\usepackage{amsfonts}
\usepackage{amssymb}
\usepackage{graphicx}%
\providecommand{\U}[1]{\protect\rule{.1in}{.1in}}
\newtheorem{theorem}{Theorem}

\newtheorem{remark}[theorem]{Remark}

\begin{document}

\title{The Mass Spectrum of Neutrinos}
\author{E. Capelas de Oliveira\thanks{capelas@ime.unicamp.br}, W. A. Rodrigues
Jr.\thanks{walrod@ime.unicamp.br\ or walrod@mpc.com.br} and J. Vaz
Jr.\thanks{vaz@ime.unicamp.br}\\$\hspace{-0.1cm}$Institute of Mathematics, Statistics and Scientific Computation\\IMECC-UNICAMP\\13083-859 Campinas, SP, Brazil}
\maketitle

\begin{abstract}
In a previous paper we showed that Weyl equation possess superluminal
solutions and moreover we showed that those\ solutions that are\ eigenstates
of the parity operator seem to describe a coupled pair of a monopole
anti-monopole system. This result suggests to look for a solution of Maxwell
equation $\boldsymbol{\partial}F^{\infty}=-\mathtt{g}\mathcal{J}$ with a
current $\mathcal{J}$ as source and such that the Lorentz force on the current
is null. We first identify a solution where $\mathcal{J=\gamma}^{5}%
J_{m\emph{\ }}$is a spacelike field (even if $F$ is not a superluminal
solution of the homogeneous Maxwell equation). More surprisingly we find that
there exists a solution $F$ of the free Maxwell $\boldsymbol{\partial}F=0$
that is equivalent to the non homogeneous equation for $F^{\infty}$. Once this
result is proved it suggests by itself to look for more general subluminal and
superluminal solutions $\mathfrak{F}$ of the free Maxwell equation equivalent
to a non homogeneous Maxwell equation for a field $\mathfrak{F}_{0}$ with a
current term as source which may be subluminal or superluminal. We exhibit one
such subluminal solution, for which the Dirac-Hestenes spinor field $\psi$
associated the electromagnetic field $\mathfrak{F}_{0}$ satisfies a Dirac
equation for a bradyonic neutrino under the \emph{ansatz} that the current is
$ce^{\lambda\gamma^{5}}\mathtt{g}\psi\gamma^{0}\tilde{\psi}$, with
$\mathtt{g}$ the quantum of magnetic charge and $\lambda$ a constant to be
determined in such a way that the auto-force be null. Together with Dirac's
quantization condition this gives a quantized mass spectrum
(Eq.(\ref{spectrum})) for the neutrinos, with the masses of the different
flavor neutrinos being of the same order of magnitude (Eq.(\ref{mv})) which is
in accord with recent experimental findings. As a last surprise we show that
the mass spectrum found in the previous case continues to hold if the current
is taken spacelike, i.e., $ce^{\lambda\gamma^{5}}\mathtt{g}\psi_{>}\gamma
^{3}\tilde{\psi}_{>}$ with $\psi_{>}$, in this case, satisfying a tachyonic
Dirac-Hestenes equation.

\end{abstract}

\section{Maxwell Equation and Neutrinos}

In a previous note we recalled that Weyl equation satisfied by a Weyl spinor
field that is an eigenvector of the parity operator seems to describe a couple
monopole anti-monopole system (CMAMS) propagating together. If this is indeed
the case than the coupled system must be solution of a Maxwell equation with a
dipolar magnetic current and such that in order for the CMAMS to be stable the
resulting Lorentz force on the CMAMS (at least classically) is null. We use
the same notations and conventions as in \cite{rodcap2007,orv2011}. All
phenomena is supposed to be well described in Minkowski spacetime containing a
preferred inertial frame eliminating causal paradoxes and breaking active
Lorentz invariance \cite{orv2011}. Then, the Maxwell equation describing the
system we are interested is%
\begin{equation}
\boldsymbol{\partial}F^{\infty}=-\mathtt{g}\gamma^{5}J_{m}=\mathtt{g}\star
J_{m}\label{1}%
\end{equation}
where \texttt{g}$\mathbb{\in R}$ denotes magnetic charge, $F^{\infty}\in\sec%
{\textstyle\bigwedge\nolimits^{2}}
T^{\ast}M\hookrightarrow\sec\mathcal{C\ell(}M,g)$, $J_{m}\in\sec%
{\textstyle\bigwedge\nolimits^{1}}
T^{\ast}M\in\sec\mathcal{C\ell(}M,g)$. Moreover, $\boldsymbol{\partial}$ is
the Dirac operator acting on sections of the Clifford bundle of differential
forms $\mathcal{C\ell(}M,g)$ and $\star$ stands for the Hodge star operator.
Now, the Lorentz force acting on $J_{m}$ is $\boldsymbol{F}=J_{m}%
\lrcorner\star F^{\infty}$ and we want that%
\begin{equation}
\boldsymbol{F}=J_{m}\lrcorner\star F^{\infty}=0.\label{2}%
\end{equation}
Let $\{\boldsymbol{e}_{\mu}=\partial/\partial\mathtt{x}^{\mu}\}$ be an
orthonormal basis for $TM$ where $(\mathtt{x}^{0},\mathtt{x}^{1}%
,\mathtt{x}^{2},\mathtt{x}^{3})=(\mathtt{t,x,y,z})$ are coordinates in the
Einstein-Lorentz-Poincar\'{e} gauge naturally adapted to the inertial frame
$\boldsymbol{e}_{0}=\frac{\partial}{\partial\mathtt{t}}$. Let $\{\gamma^{\mu
}\}$ be a basis for $T^{\ast}M$ dual to the basis $\{\boldsymbol{e}_{\mu}\}.$
Write $J_{m}=\rho_{m}\boldsymbol{\gamma}^{0}-j_{mi}\boldsymbol{\gamma}^{i}$
and introduce engineering notation (using the well known fact that the even
sub-bundle $\mathcal{C\ell}^{0}\mathcal{(}M,g)$ of $\mathcal{C\ell(}M,g)$ is
isomorphic to the Pauli bundle) writing%
\begin{equation}
F^{\infty}=\frac{1}{2}F_{\mu\nu}\boldsymbol{\gamma}^{\mu}\boldsymbol{\gamma
}^{\nu}:=\mathbf{E}_{\infty}\mathbf{+}\gamma^{5}\mathbf{B}_{\infty}%
=\mathbf{E}_{\infty}\mathbf{-iB}_{\infty}\label{3}%
\end{equation}
with%
\begin{align}
\mathbf{E}_{\infty} &  \mathbf{=}%
{\textstyle\sum\nolimits_{i=1}^{3}}
F_{0i}\boldsymbol{\gamma}^{0}\boldsymbol{\gamma}^{i}=%
{\textstyle\sum\nolimits_{i=1}^{3}}
F_{0i}\boldsymbol{\sigma}_{i},\nonumber\\
\mathbf{B}_{\infty} &  \mathbf{=}\frac{1}{2}%
{\textstyle\sum\nolimits_{i,j=1}^{3}}
F_{ij}\boldsymbol{\gamma}^{i}\boldsymbol{\gamma}^{j}=-\frac{1}{2}%
{\textstyle\sum\nolimits_{i,j=1}^{3}}
F_{ij}\boldsymbol{\gamma\boldsymbol{\sigma}_{i}\sigma}_{j}\nonumber\\
\star F^{\infty} &  =\mathbf{i}F\mathbf{^{\infty}=B}_{\infty}+\mathbf{iE}%
_{\infty}.\label{4}%
\end{align}
Projecting $\boldsymbol{F}$ in Eq.(\ref{2}) in the Pauli bundle selected by
the inertial reference frame $\boldsymbol{e}_{0}$ we get\footnote{The
Euclidian scalar product will be denoted by the symbol \textquotedblleft%
$\bullet$\textquotedblright.}%
\begin{equation}
\boldsymbol{F\gamma}^{0}=\mathbf{j}_{m}\bullet\mathbf{B}_{\infty}%
\mathbf{-(}\rho_{m}\mathbf{B}_{\infty}+\mathbf{\mathbf{j}_{m}\times E}%
_{\infty}\mathbf{),}\label{5}%
\end{equation}
and we immediately realize that we can have $\boldsymbol{F}=0$ if $J_{m}$ is a
spacelike or a timelike current. For the first case it exists an inertial
reference frame where the current is \emph{transcendent\footnote{We take
$\mathbf{e}_{\mu}=\partial/\partial\mathtt{x}^{0}=\partial/\partial\mathtt{t}$
as that system and say that the CMAMS is transcendent.}}, i.e., where
$\rho_{m}=0,$ $\mathbf{j}_{m}\neq0$, but this is not, of course true in any
reference frame. Thus to have $\boldsymbol{F}=0$ we need in the general case
\begin{equation}
\mathbf{\mathbf{j}_{m}}=b\mathbf{E}_{\infty},\text{ \ \ }\mathbf{E}_{\infty
}\mathbf{\bullet B}_{\infty}=0,\label{6}%
\end{equation}
which force us to take $\mathbf{B}_{\infty}=0$. This means that in
Eq.(\ref{1}) we must take
\begin{equation}
F^{\infty}=%
{\textstyle\sum\nolimits_{i=1}^{3}}
F_{0i}\boldsymbol{\gamma}^{0}\boldsymbol{\gamma}^{i}=%
{\textstyle\sum\nolimits_{i=1}^{3}}
F_{0i}\boldsymbol{\sigma}_{i}.\label{7}%
\end{equation}
We now write Eq.(\ref{1}) in the Pauli algebra associated to the inertial
frame for the case of a spacelike current and in the \ reference frame where
it is transcendent. We have\ using that $\rho_{m}=0$ and $b=1$,
\begin{align}
\mathbf{j}_{m} &  =J_{m}\boldsymbol{\gamma}_{0}=j_{mi}\boldsymbol{\gamma}%
_{i}\boldsymbol{\gamma}_{0}=b\mathbf{E}_{\infty}=b%
{\textstyle\sum\nolimits_{i=1}^{3}}
F_{0i}\boldsymbol{\sigma}_{i}\nonumber\\
&  =b%
{\textstyle\sum\nolimits_{i=1}^{3}}
F_{0i}\boldsymbol{\gamma}^{0}\boldsymbol{\gamma}^{i},\label{8}%
\end{align}
and thus%
\begin{equation}
J_{m}=(%
{\textstyle\sum\nolimits_{i=1}^{3}}
F_{0i}\boldsymbol{\gamma}^{0}\boldsymbol{\gamma}^{i}\boldsymbol{)\gamma}%
^{0}=F^{\infty}\boldsymbol{\gamma}^{0}.\label{9}%
\end{equation}
Then Eq.(\ref{1}) becomes
\begin{equation}
\boldsymbol{\partial}F^{\infty}=-\mathtt{g}\gamma^{5}F^{\infty}%
\boldsymbol{\gamma}^{0}.\label{10}%
\end{equation}
It is quite clear that in the reference frame where the current is
transcendental its support for each instant of time is $\mathbb{R}^{3}$ and in
that frame we suppose that $F^{\infty}$ is static, i.e., it is not a function
of $t$. Under these conditions\ we can write Eq.(\ref{10}) as
\begin{equation}
-\boldsymbol{\gamma}^{i}\boldsymbol{\gamma}^{0}\partial_{i}%
{\textstyle\sum\nolimits_{i=1}^{3}}
F_{0i}\boldsymbol{\sigma}_{i}=\mathtt{g}\mathbf{i}%
{\textstyle\sum\nolimits_{i=1}^{3}}
F_{0i}\boldsymbol{\sigma}_{i}\label{11}%
\end{equation}
or yet,%
\begin{equation}
\nabla\mathbf{E}_{\infty}=\mathtt{g}\mathbf{iE}_{\infty}\mathbf{.}\label{12}%
\end{equation}

Finally observing that the total electric charge density involved in our
problem is always null we must have $\nabla\bullet\mathbf{E}_{\infty}=0$ and
thus Eq.(\ref{12}) can be written%
\begin{equation}
-\mathbf{i(}\nabla\wedge\mathbf{E}_{\infty}\mathbf{)=}2\mathtt{g}%
\mathbf{E}_{\infty}. \label{13}%
\end{equation}
Taking into account that $-\mathbf{i(}\nabla\wedge\mathbf{E}_{\infty
}\mathbf{):=}\nabla\times\mathbf{E}_{\infty}$ we end with the surprising
result that the CMAMS electromagnetic field satisfies a force free equation,
i.e.,
\begin{equation}
\nabla\times\mathbf{E}_{\infty}=2\mathtt{g}\mathbf{E}_{\infty}. \label{14}%
\end{equation}
Applying the $\nabla\times$ to both members of that equation we get taking
into account that $\nabla\bullet\mathbf{E}_{\infty}=0$ that%
\begin{equation}
\nabla^{2}\mathbf{E}_{\infty}+\mathtt{g}^{2}\mathbf{E}_{\infty}=0. \label{15}%
\end{equation}
If we suppose that the possible values of the magnetic charge obeys Dirac
quantization condition, i.e. $e\mathtt{g}=n/2$, $n\in\mathbb{Z}$, then it
results that the possible values of $\mathtt{g}^{2}$ is also quantized.

Our surprises did not end yet. Indeed, we now show that a \emph{free}
electromagnetic field can simulate Eq.(\ref{1}). In order to show this result
write
\begin{align}
F  &  =F^{\infty}\exp(\boldsymbol{\gamma}^{5}m\mathtt{t})\nonumber\\
&  =\mathbf{E}_{\infty}\cos m\mathtt{t}-\mathbf{iE}_{\infty}\sin
m\mathtt{t}\nonumber\\
&  =\mathbf{E+iB.} \label{16}%
\end{align}
Recalling again that the density of magnetic charge for a transcendent CMAMS
is null we immediately realize that%
\begin{gather*}
\nabla\bullet\mathbf{E}=\nabla\bullet\mathbf{B}=0,\\
\nabla\times\mathbf{E}+\frac{\partial}{\partial\mathtt{t}}\mathbf{B}=0,\text{
\ \ }\nabla\times\mathbf{B}-\frac{\partial}{\partial\mathtt{t}}\mathbf{E}=0
\end{gather*}
i.e., $\boldsymbol{\partial}F=0.$

Then, taking into account that $F^{\infty}$ is independent of time we can
write
\begin{equation}
\boldsymbol{\partial}(F^{\infty}\cos m\mathtt{t})=+\boldsymbol{\gamma}%
^{5}\boldsymbol{\partial}(F^{\infty}\sin m\mathtt{t}), \label{21}%
\end{equation}
getting
\begin{equation}
\boldsymbol{\gamma}^{i}\partial_{i}F^{\infty}=m\boldsymbol{\gamma}%
^{5}\boldsymbol{\gamma}^{0}F^{\infty}, \label{23}%
\end{equation}
and since $\boldsymbol{\gamma}^{0}\wedge F^{\infty}=0$ we have that
$\boldsymbol{\gamma}^{0}F^{\infty}=\boldsymbol{\gamma}^{0}\lrcorner F^{\infty
}+\boldsymbol{\gamma}^{0}\wedge F^{\infty}=-F^{\infty}\llcorner
\boldsymbol{\gamma}^{0}=-F^{\infty}\boldsymbol{\gamma}^{0}$ and Eq.(\ref{23})
becomes
\begin{subequations}
\label{24}%
\begin{gather}
\boldsymbol{\partial}F^{\infty}=-m\boldsymbol{\gamma}^{5}F^{\infty
}\mathbf{\gamma}^{0},\label{a}\\
\boldsymbol{\partial}^{2}F^{\infty}-m^{2}F^{\infty}=0. \label{b}%
\end{gather}
Eq.(\ref{a}) is exactly Eq.(\ref{1}), the Maxwell equation supposed to
describe a CMAMS with a \emph{spacelike current} (by Eq.(\ref{a})) once we
take into account (in appropriate units) $m$ to be proportional to
$\mathtt{g}$ , i.e.,
\end{subequations}
\begin{equation}
m=\mathfrak{k}\mathtt{g.} \label{24a}%
\end{equation}

\begin{remark}
In what follows we will see that in our theory the proportional factor
$\mathfrak{k}$ can take only a discrete set of values thus determining a
discrete set of neutrinos mass values.
\end{remark}

\begin{remark}
Another very interesting fact is the following one. Let $\{e_{\mu}%
=\partial/\partial x^{\mu}\}$ be a basis for $TM$ where $(x^{0},x^{1}%
,x^{2},x^{3})=(t,x,y,z)$ are coordinates in the Einstein-Lorentz-Poincar\'{e}
gauge naturally adapted to the inertial frame $e_{0}=\partial/\partial x^{\mu
}$ moving with respect to $\boldsymbol{e}_{0}$, i.e.,
\begin{equation}
e_{0}=\frac{1}{\sqrt{1-V^{2}}}\boldsymbol{e}_{0}+\frac{V}{\sqrt{1-V^{2}}%
}\boldsymbol{e}_{3}. \label{24A}%
\end{equation}
In that frame we have that
\begin{align}
F  &  =F^{\infty}\exp[\boldsymbol{\gamma}^{5}(\frac{mt}{\sqrt{1-V^{2}}}%
-\frac{mVz}{\sqrt{1-V^{2}}})]\nonumber\\
&  =F^{\infty}\exp[\boldsymbol{\gamma}^{5}(\omega t-kz)], \label{24b}%
\end{align}
and of course $\omega^{2}-k^{2}=m^{2}$, i.e., we have a subluminal free
electromagnetic field configuration. But, of course the magnetic current
associated to $F^{\infty}$ continues to be superluminal.
\end{remark}

This result suggests us to look in an arbitrary basis\footnote{The dual basis
of $\{e_{\mu}\}$ is denoted in what follows by $\{\gamma^{\mu}\}$.} $\{e_{\mu
}=\partial/\partial x^{\mu}\}$ of $TM$ (where $(x^{0},x^{1},x^{2}%
,x^{3})=(t,x,y,z)$ are coordinates in the Einstein-Lorentz-Poincar\'{e} gauge
naturally adapted to the inertial frame $e_{0}=\partial/\partial x^{\mu}$) for
solutions of the free Maxwell equation%
\begin{equation}
\boldsymbol{\partial}\mathfrak{F}=0, \label{free}%
\end{equation}
describing subluminal and superluminal neutrinos, i.e., such that
Eq.(\ref{free}) be equivalent to another nonhomogeneous Maxwell equation for
$\mathfrak{F}_{0}$ with a \emph{subluminal} or \textit{superluminal}
\emph{magnetic current source}, but in both cases the Lorentz force being
$J_{m}\lrcorner\mathfrak{F}_{0}=0$.

To simplify our task we suppose that the solution we are looking for is
propagating in the $z$-direction of the inertial frame $e_{0}$ at speed $v$
associated with the phase of the duality rotation. By this statement we means
that our solution must have the form
\begin{equation}
\mathfrak{F}=\mathfrak{F}_{0}\exp[-\gamma^{5}(\omega t-kz)]=\mathfrak{F}%
_{0}e^{-\gamma^{5}\chi}\label{f1}%
\end{equation}
with $\kappa=\omega\gamma^{0}-k\gamma^{3}$ and $\chi=\omega t-kz$. We have two
cases that need to be analyzed: (i) $\kappa_{<}^{2}=\omega_{<}^{2}-k_{<}%
^{2}=m_{<}^{2}$ and (ii) $\kappa_{>}^{2}=\omega_{>}^{2}-k_{>}^{2}=-m_{>}^{2}$.
For both cases (i) and (ii) we write respectively\footnote{The notation even
if confused at a first sight will becomes obvious in a while.}
\begin{equation}
\mathfrak{F}^{<}=\mathfrak{F}_{0}^{>}e^{-\gamma^{5}\chi_{<}},\text{
\ \ \ \ }\mathfrak{F}^{>}=\mathfrak{F}_{0}^{\,<}e^{-\gamma^{5}\chi_{>}%
}.\label{f22}%
\end{equation}
\ \ \ 

The equivalent Maxwell equation for $\mathfrak{F}_{0}^{>}$ ($\mathfrak{F}%
_{0}^{<}$) has a superluminal (subluminal) magnetic current $\kappa
_{<}\mathfrak{F}_{0}^{>}=\kappa_{<}\lrcorner\mathfrak{F}_{0}^{>}$ ($\kappa
_{>}\mathfrak{F}_{0}^{<}=\kappa_{>}\lrcorner\mathfrak{F}_{0}^{<}$).
Indeed,\ for $\mathfrak{F}_{0}^{>}$ we have from Eq.(\ref{f1}) that%
\begin{equation}
\partial\mathfrak{F}_{0}^{>}=-\gamma^{5}\kappa_{<}\mathfrak{F}_{0}^{>},
\label{f11}%
\end{equation}
\ \ which means that $\mathfrak{F}_{0}^{>}$ must also satisfy the
\emph{tachyonic }Klein-Gordon equation,
\begin{equation}
\partial^{2}\mathfrak{F}_{0}^{>}-m_{<}^{2}\mathfrak{F}_{0}^{>}=0. \label{f3}%
\end{equation}

For the case (ii) the equivalent equation for $\mathfrak{F}_{0}^{<}$ \ is again%

\begin{equation}
\partial\mathfrak{F}_{0}^{<}=-\gamma^{5}\kappa_{>}\mathfrak{F}_{0}^{<}
\label{f4}%
\end{equation}
but now it has a subluminal magnetic current $\kappa_{>}\mathfrak{F}_{0}%
^{<}=\kappa_{>}\lrcorner\mathfrak{F}_{0}^{<}$ since Eq.(\ref{f4}) gives a
\emph{bradyonic} Klein-Gordon equation, i.e.,%
\begin{equation}
\partial^{2}\mathfrak{F}_{0}^{<}+m_{>}^{2}\mathfrak{F}_{0}^{<}=0. \label{f5}%
\end{equation}

To obtain a solution $\mathfrak{F}_{0}^{<}$ satisfying Eqs.(\ref{f4}) and
(\ref{f5}), we introduce a \emph{Hertz potential }$\Pi^{>}\in\sec%
{\textstyle\bigwedge\nolimits^{2}}
T^{\ast}M\in\sec\mathcal{C\ell(}M,g)$ satisfying $\square\Pi^{>}=0$ such that
the electromagnetic potential is $A^{>}=-\delta\Pi^{>}$.\ Thus since $\delta
A^{>}=0$ and $\mathfrak{F}=dA^{>}$ it follows that $\boldsymbol{\partial
}\mathfrak{F}^{>}=0$. The Hertz potential satisfies $\square\Pi^{>}=0$ and we
look for solutions of the form%
\begin{equation}
\Pi^{>}=\Phi_{>}(x,y)e^{\gamma^{5}(\omega_{>}t-k_{>}z)}\mathbf{B,} \label{25}%
\end{equation}
\medskip where $\mathbf{B}$ is a constant $2$-form. It follows that
\begin{gather}
-\omega_{>}^{2}\Phi_{>}(x,y)\exp[\gamma^{5}(\omega_{>}t-k_{>}z)]\mathbf{B}%
\nonumber\\
-\nabla_{2}^{2}\Phi_{>}(x,y)\exp[\gamma^{5}(\omega_{>}t-k_{>}z)]\mathbf{B}%
\nonumber\\
+k_{>}^{2}\Phi_{>}(x,y)\exp[\gamma^{5}(\omega_{>}t-k_{>}z)]\mathbf{B}=0.
\label{28}%
\end{gather}
Then, if we choose%
\begin{equation}
\nabla_{2}^{2}\Phi_{>}(x,y)=m_{>}^{2}\Phi_{>}(x,y) \label{29}%
\end{equation}
where $\nabla_{2}^{2}$ means the $2$-dimensional Laplacian we get the desired
dispersion relation, i.e.,%
\begin{equation}
\omega_{>}^{2}-k_{>}^{2}=-m_{>}^{2}, \label{29b}%
\end{equation}
and $\mathfrak{F}_{0}^{<}$ satisfies Eq.(\ref{f5}), a bradyonic Klein-Gordon
equation. \medskip

On the other hand to obtain a solution $\mathfrak{F}_{0}^{>}$ satisfying
Eq.(\ref{f3}), we introduce a \emph{Hertz potential }$\Pi^{<}=\Phi
_{<}(x,y)e^{\gamma^{5}(\omega_{>}t-k_{>}z)}\mathbf{B}$, repeat the above steps
but impose that $\nabla_{2}^{2}\Phi_{<}(x,y)=-m_{<}^{2}\Phi_{<}(x,y)$.

Thus, we have proved as stated above that the free Maxwell equation has
solutions with a bradyonic dispersion relation, which is equivalent to another
Maxwell equation with a tachyonic magnetic current and also has a solution
with tachyonic dispersion relation which is equivalent to another Maxwell
equation with a bradyonic magnetic current. We now analyze the case of the
bradyonic current associated to a coupled monopole anti-monopole pair (i.e.,
the subluminal neutrino). The analysis for the other case is similar, and
conduces to the same mass spectrum.

\section{Neutrino Mass Spectrum}

It remains to answer a crucial question. What kind of spinor field is
associated with a subluminal ( or a superluminal) field $\mathfrak{F}_{0}%
$\footnote{To simplify notation in this section $\mathfrak{F}_{0}^{<}$ is
denoted $\mathfrak{F}_{0}$.}?

The answer is that since $\mathfrak{F}_{0}^{2}\neq0$, we must associated to
$\mathfrak{F}_{0}$ a Dirac-Hestenes spinor field \cite{mr2004} whose
representative in the spinorial frame associated to $e_{0}$ is $\psi\in
\sec\mathcal{C\ell}^{0}\mathcal{(}M,g)$. Moreover, we assume that $\psi
\tilde{\psi}\neq0$, which means that $\psi$ has the following decomposition%
\begin{equation}
\psi=\varrho^{1/2}e^{\beta\gamma^{5}/2}R, \label{30}%
\end{equation}
where $\rho$ and $\beta$ are scalar functions and for each spacetime point
$x$, $R(x)\in Spin_{1,3}^{e}$ since we want \ to write for the field of each
neutrino flavor\footnote{That it is always possible to write $\mathfrak{F}%
_{0}$ as in Eq.(\ref{30a}) is proved in \cite{vr1993}.} in Gaussian units%
\begin{equation}
\mathfrak{F}_{0}=\frac{2\pi\mathtt{e}\hbar}{\mathtt{m}c}\psi\gamma^{2}%
\gamma^{1}\tilde{\psi}, \label{30a}%
\end{equation}
where \texttt{e} is the electronic charge and \texttt{m} is a mass parameter
to be determined experimentally. We return to Eq.(\ref{f11}) which we write
from now on in Gaussian units as%
\begin{equation}
\partial\mathfrak{F}_{0}=\frac{4\pi}{c}\mathcal{J}. \label{30b}%
\end{equation}

Under the above conditions\ we can show \cite{vr1993,rv1998} that
Eq.(\ref{30b}) is equivalent (once we impose a very reasonable constraint to
match the number of degrees of freedom $\mathfrak{F}_{0}$ with the ones of
$\psi$) to the following Dirac-like equation,%
\begin{gather}
\boldsymbol{\partial}\psi\gamma^{2}\gamma^{1}=\frac{\Lambda c}{\hbar}%
\psi\gamma_{0}e^{\beta\gamma^{5}}\nonumber\\
+\gamma^{5}\frac{Kc}{\hbar}\psi\gamma_{0}e^{\beta\gamma^{5}}+\frac
{e^{\beta\gamma^{5}}}{\rho}\left(  \frac{\mathtt{m}}{\mathtt{e}\hbar}\right)
\mathcal{J}\psi, \label{31}%
\end{gather}
where the scalar functions $\Lambda$ and $K$ are given by%

\begin{equation}
\Lambda=\Omega\cdot S,\text{ \ \ \ }K=\Omega\cdot(\gamma^{5}S), \label{32}%
\end{equation}
with $S=\frac{1}{2}R\gamma^{2}\gamma^{1}\tilde{R}$, $\Omega=v^{\mu}\Omega
_{\mu}$, $v^{\mu}=(R\gamma^{0}\tilde{R})\cdot\gamma^{\mu}$ and
\begin{equation}
\Omega_{\mu}=2(\partial_{\mu}R)\tilde{R}. \label{33}%
\end{equation}

In \cite{rv1998} by taking $K=0$ and $\Lambda=\mathtt{m}$ (the mass of the
electron), $J_{m}=c\mathtt{g}\phi\gamma^{0}\tilde{\phi}$ (with $\phi
=e^{-\beta\gamma^{5}}\psi)$ and considering the muon and the tau as excited
states of electron (containing pairs of monopole anti-monopole), we determine
the spectrum of the lepton family (giving the electron mass) using Dirac's
quantization condition with $\mathtt{e}=e/3$ as minimum charge\ ($e$ being the
electronic charge). We found the values $105.5$ MeV and $1785$ MeV for the
muon and tau masses, respectively.

Here instead we take $\Lambda=0$ and determine $K\ $by imposing the
\emph{ansatz}%
\begin{equation}
\mathcal{J}=e^{\lambda\gamma^{5}}c\mathtt{g}\psi\gamma^{0}\tilde{\psi},
\label{ansatz}%
\end{equation}
where $\lambda$ is to be determined in such a way that the auto-force
$\mathcal{J\lrcorner}\mathfrak{F}_{0}=0$. At once we find that this implies%
\begin{equation}
\lambda=\beta\pm n\pi\label{ans1}%
\end{equation}
and without loosing generality we take in what follows $\lambda=\beta$.

Then, after some algebra we get
\begin{equation}
\boldsymbol{\partial}\psi\gamma^{2}\gamma^{1}=\frac{m_{1}c}{\hbar}\psi
\gamma_{0}+\gamma^{5}\frac{m_{2}c}{\hbar}\psi\gamma_{0} \label{35}%
\end{equation}
where
\begin{subequations}
\label{36}%
\begin{align}
m_{1}  &  =\left(  K\sin\beta+\frac{\mathtt{mg}}{\mathtt{e}}\cos\beta\right)
,\label{m1}\\
m_{2}  &  =\left(  K\cos\beta+\frac{\mathtt{mg}}{\mathtt{e}}\sin\beta\right)
. \label{m2}%
\end{align}

We now impose that $m_{1}=0$, which implies that%
\end{subequations}
\begin{equation}
\tan\beta=-\frac{1}{K}\left(  \frac{\mathtt{mg}}{\mathtt{e}}\right)  .
\label{c1}%
\end{equation}

Then%
\begin{equation}
m_{2}=\frac{K^{2}-\left(  \frac{\mathtt{mg}}{\mathtt{e}}\right)  ^{2}}%
{\sqrt{K^{2}+\left(  \frac{\mathtt{mg}}{\mathtt{e}}\right)  ^{2}}}. \label{c2}%
\end{equation}
We take again the value $\mathtt{e}=e/3$, and use Dirac's quantization
condition ($\alpha$ is the fine structure constant)
\begin{equation}
\mathtt{g}e/3\hbar c=\frac{n}{2}. \label{dl}%
\end{equation}
With the use of Eq.(\ref{dl}), the Eq.(\ref{c2}) becomes ($m_{\nu_{n}}=m_{2}$)%

\begin{equation}
m_{\nu_{n}}=\frac{K^{2}-\frac{9\mathtt{m}^{2}}{4\alpha^{2}}n^{2}}{\sqrt
{K^{2}+\frac{9\mathtt{m}^{2}}{4\alpha^{2}}n^{2}}} \label{SPEC}%
\end{equation}

Now, if the neutrino are massive particles then the allowed values of
$n\in\mathbb{Z}$ in Eq.(\ref{SPEC}) must satisfy $K\geq3\mathtt{m}n/2\alpha$
and we write%
\begin{equation}
K=3\mathtt{m}N/2\alpha,\label{cond}%
\end{equation}
for some constant $N$, which does nt need to be an integer. Then the spectrum
is%
\begin{equation}%
\begin{tabular}
[c]{|l|}\hline
$m_{\nu_{n}}=\frac{3\mathtt{m}}{2\alpha}\frac{(N^{2}-n^{2})}{\sqrt{N^{2}%
+n^{2}}}$\\\hline
\end{tabular}
\ \ \label{spectrum}%
\end{equation}
So, our formula for the spectrum depends on two parameters $\mathtt{m}$ and
$N$, but what is really important to emphasize here is that the allowed masses
for the possible neutrino mass flavors are all very near, which seems to be in
agreement with recent experimental data \cite{tho}. Since there exists three
different flavors we take provisorily, in order to get an estimative of the
masses $N=3$. Under the constraint given in \cite{tho} that the masses of the
neutrinos is less than $0.28$ eV we get $\mathtt{m}=1.97$ $10^{-4}$ eV and the
neutrino masses results%
\begin{equation}
m_{\nu_{0}}=0.12\text{ eV, \ \ }m_{\nu_{1}}=0.10\text{ eV,...,}m_{\nu_{2}%
}=0.056\text{ eV.}\label{mv}%
\end{equation}

\begin{remark}
We observe that with the values given in \emph{Eq.(\ref{mv})} we have
$m_{\nu_{0}}^{2}-m_{\nu_{1}}^{2}=4.4$ $10^{-5}$ and $m_{\nu_{1}}^{2}%
-m_{\nu_{2}}^{2}=6.86$ $10^{-3}$, $m_{\nu_{0}}^{2}-m_{\nu_{2}}^{2}=16.46$
$10^{-3}$ which altough not equal to published data \cite{naka} are of the
same order of magnitude, but more cannot be inferred here because the mass
formula is very sensitive to the values of the parameters.
\end{remark}

Now, returning to Eq.(\ref{35}) we get%

\begin{equation}
\boldsymbol{\partial}\psi\gamma^{2}\gamma^{1}=\gamma^{5}\frac{m_{\nu_{n}}%
c}{\hbar}\psi\gamma_{0}. \label{SUPD}%
\end{equation}

This is a Dirac equation for a bradyonic neutrino since the following
Klein-Gordon equation holds ($\psi=\psi_{<}$)%

\begin{equation}
\boldsymbol{\partial}^{2}\psi_{<}+\left(  \frac{m_{\nu_{n}}c}{\hbar}\right)
^{2}\psi_{<}=0. \label{37}%
\end{equation}

\begin{remark}
Note that if neutrinos are tachyonic particles\ satisfying
\begin{equation}
\boldsymbol{\partial}^{2}\psi_{>}-\left(  \frac{m_{\nu_{n}}c}{\hbar}\right)
^{2}\psi_{>}=0, \label{38}%
\end{equation}
we may find a mass spectrum identical to the one given by
\emph{Eq.(\ref{spectrum}) }following steps analogous to the ones that lead to
Eq.(\ref{35}).\emph{ }We choose
\[
\mathfrak{F}_{0}^{>}=\frac{2\pi\mathtt{e}\hbar}{\mathtt{m}c}\psi_{>}\gamma
^{0}\gamma^{3}\tilde{\psi}_{>},
\]
\emph{use }$m_{1}=0$\emph{,} but \ now take $\mathcal{J}=ce^{\lambda\gamma
^{5}}\mathtt{g}\psi_{>}\gamma^{3}\tilde{\psi}_{>}$ \emph{(}a superluminal
current\emph{). We arrive to} the following corresponding tachyonic
Dirac-Hestenes\ equation for $\psi_{>}$\emph{:}%
\begin{equation}
\boldsymbol{\partial}\psi_{>}=\gamma^{5}\frac{m_{\nu_{n}}c}{\hbar}\psi
_{>}\gamma^{0}. \label{39}%
\end{equation}
Thus $\psi_{>}$ satisfies \emph{Eq.(\ref{38}).}
\end{remark}

We observe that the mass spectrum found for the neutrinos is compatible (with
reasonable choice of the parameters of the theory) with the recent
experimental results \cite{tho} showing that the sum of the masses of the
known flavor neutrinos is less that $0.28$ eV ($95\%$ CL), for Eq.(\ref{SPEC})
show that the masses for the different flavours do not differ very much, the
lightest neutrino having a mass less than $0.056$ eV. However under these
conditions our mass spectrum is not compatible with the data of the OPERA
experiment that (if reliable) gives\ a tachyonic muonic neutrino mass of $120$
MeV. Finally, it is necessary to mention that the idea that monopoles are
excited states of the neutrino was first proposed by Lochak \cite{lo}, but he
never obtained results as the ones described above.

\end{document}